%% file: shinkaikandaebisu.tex
\documentclass[twocolumn]{aastex6}

\begin{document}

\title{Gravitational Waves from Merging Intermediate-mass Black Holes : \\
II Event Rates at Ground-based Detectors
}

\author{Hisa-aki Shinkai\altaffilmark{1}}
\affil{Dept. of Information Science and Technology, Osaka Institute of Technology, Kitayama 1-79-1, Hirakata City, Osaka 573-0196, Japan}
\author{Nobuyuki Kanda\altaffilmark{2}}
\affil{Dept. of Physics, Osaka City University, 
Sugimoto 3-3-138, Sumiyoshi, Osaka City, Osaka 558-8585, Japan}
\and 
\author{Toshikazu Ebisuzaki\altaffilmark{3}}
\affil{Computational Astrophysics Laboratory, Institute of Physical and Chemical Research (RIKEN), Hirosawa 2-1, Wako City, Saitama 351-0198, Japan}

\altaffiltext{1}{hisaaki.shinkai@oit.ac.jp}
\altaffiltext{2}{kanda@sci.osaka-cu.ac.jp}
\altaffiltext{3}{ebisu@riken.jp}

\begin{abstract}
Based on a dynamical formation model of a supermassive black hole (SMBH),  
we estimate the expected observational profile of gravitational wave at ground-based detectors, such as KAGRA or 
advanced LIGO/VIRGO.
Noting that the second generation of detectors have enough sensitivity from 10 Hz and up  
(especially with KAGRA owing to its location at less seismic noise),  
we are able to detect the ring-down gravitational wave of a BH with the mass $M < 2\times 10^3 M_\odot $. 
This enables us to check the sequence of BH mergers to SMBHs via intermediate-mass BHs. 
We estimate the number density of galaxies from the halo formation model and estimate the number of 
BH mergers from the giant molecular cloud model assuming hierarchical growth of merged cores. 
At the designed KAGRA (and/or advanced LIGO/VIRGO), we find that 
the BH merger of its total mass $M\sim 60M_\odot$ is at the peak of the expected mass distribution.  
With its signal-to-noise ratio $\rho=10 (30)$, we estimate the event rate $R \sim 200 (20)$ per year in the most optimistic case, and we also find that BH mergers in the range $M < 150 M_\odot$ are $R>1$ per year for $\rho=10$. 
Thus, if we observe a BH with more than $100 M_\odot$ in future gravitational-wave observations, our model naturally explains its source. 
\footnote{Published as Astrophysical Journal  {\bf 835} (2017), 276. \\
\url{http://dx.doi.org/10.3847/1538-4357/835/2/276}\\ arXiv:1610.09505v3. }
\end{abstract}


\section{Introduction}
\subsection{Era of Gravitational-wave Astronomy}
\label{sec1}
The direct detections of gravitational waves were announced by the advanced LIGO group in 2016
\citep{GW150914,GW151226}, and we are at the opening era of ``gravitational-wave astronomy".
The LIGO group reported two events (GW150914, GW151226) and one transient event (LVT151012), 
all three of which are regarded as the events of coalescence of binary black holes (BBHs). 

The first event (GW150914) was the merger of BHs of the masses
$36.2^{+ 5.2}_{-3.8} M_\odot$ and $29.1^{+ 3.7}_{-4.4} M_\odot$, which turned into a single BH of 
$62.3^{+ 3.7}_{-3.1} M_\odot$ with spin $a=0.68^{+ 0.05}_{-0.06}$, 
which shows that the energy radiation rate is 4.6\% of the total mass. 
The event occurred at redshift $z=0.09^{+0.03}_{-0.04}$, and was 
detected with signal-to-noise ratio (S/N) $\rho=23.7$. 
The second event (GW151226) was the merger of BHs with 
$14.2^{+ 8.3}_{-3.7} M_\odot$ and $7.5^{+ 2.3}_{-2.3} M_\odot$, which turned into a single BH of 
$20.8^{+ 6.1}_{-1.7} M_\odot$ with spin $a=0.74^{+ 0.06}_{-0.06}$, 
which shows that the energy radiation rate is 4.1\% of the total mass. 
The event occurred at redshift $z=0.09^{+0.03}_{-0.04}$, and was detected with $\rho=13.0$ 
(these numbers were taken from \cite{GW151226b}).

These announcements were not only valuable on the point of the direct detections of the gravitational wave, 
but also the first results of confirming the existence of BHs, the existence of BHs of 
this mass range, and the existence of BBHs. 
Especially, the existence of $\sim 30  M_\odot$ BHs was surprising to the community, since there were no
such observational evidences ever before.

\subsection{Possible Sources of 30 $M_\odot$ BHs}
The traditional scenarios for forming BBHs are common envelope evolution of primordial binary massive stars
\citep{Belczynski2016}, and dynamical formation in dense star clusters \citep{portegies2000}. 

One of the possible scenario is to suppose BBHs from Population III stars \citep{BondCarr1984,Belczynski2004}. 
Recently, \citet{Kinugawa2014,Kinugawa2016} predicted event rates based on this model. 
Existence of Population III stars is yet to be confirmed, but they show that 
a typical BH mass of this model is at 
$\sim 30 M_\odot$ (chirp mass $\sim 60 M_\odot$), and the event rate would be 500 yr$^{-1}$ (50 yr$^{-1}$
for $\rho\geq 20$) \citep{NakanoTanakaNakamura2015}.

Recently, \citet{FujiiTanikawaMakino2016} estimate BH mergers combining their $N$-body simulations, modeling of globular clusters, and cosmic star-cluster formation history and find that BH mass distribution has a peak at 10 $M_\odot$ and 50 $M_\odot$, and the event rate for designed LIGO is at most 85 yr$^{-1}$. 

In this article, based on the formation scenario of a supermassive BH (SMBH), 
we extend the previous model to a sequence of intermediate-mass BHs (IMBHs), and 
estimate their observational detectability at ground-based gravitational-wave detectors. 

\subsection{SMBH Runaway Path}
The formation process of an SMBH is one of the unsolved problems in galaxy evolution history. 
Many possible routes were suggested by \citet{rees1978} 
long ago, { but we still debate a plausible route. 
We do not yet know whether the first generation of BHs are of stellar-mass size or supermassive. 
See, e.g., \citet{Volonteri2012} and \citet{Haiman2013} for a review. 

One of the simplest scenarios for forming an SMBH is from the direct collapse of gas 
clouds or supermassive stars, or massive disks (e.g. \cite{Ume1993, LoebRasio1994, ShibataShapiro2002, BrommLoeb2004, Begelman2006, Begelman2008}). 
Another scenario is by accretions onto, or mergers of, the remnants of Population III stars (e.g. \cite{HaimanLoeb2001, VolonteriBegelman2010, Johnson2012, Johnson2013}).  Recent studies suggest that we can construct a formation route of SMBHs without contradicting with current observations. 

In this article, we take the third route: accumulations of BHs.  
This route was came to be believed when an IMBH   
($10^2$--$10^3 M_\odot$) was first discovered in a starburst galaxy M82 
\citep{matsumoto2001,matsushita2000}.  
So far, many IMBHs have been found in the center of galaxies 
(for a review, see, e.g. \citet{Greene2012, Yagi2012}), and also the existence of an 
IMBH of $10^4 M_\odot$ close to the Sgr A$^\ast$ has recently been reported \citep{Tsuboi2016}
(see also \citet{PZ2006,FIYM2008}). 
}

This runaway path was first proposed by \citet{ebisuzaki2001}. 
The scenario consists of three steps: 
(1) formation of IMBHs by runaway mergers of massive stars in dense star clusters
\citep{portegies2004}, 
(2) accumulations of IMBHs at the center region of a galaxy due to sinkages of clusters by dynamical friction, 
and 
(3) mergers of IMBHs by multibody interactions and gravitational radiation. 
Successive mergers of IMBHs are likely to form an SMBH with a mass of heavier than $>10^6M_\odot$. 
\citet{ebisuzaki2001} predicted that IMBH--IMBH or IMBH--SMBH merging events could be observed on the order of one per month or even one per week. 

Numerical simulations support the above first step
\citep{MarchantShapiro, portegies1999, portegies2002, portegies2004, BaumgardtMakino2003}, 
and the second step is also confirmed in a realistic mass-loss model
\citep{matsubayashi2007}, 
while the third step is not yet investigated in detail. 
The discovery of an SMBH binary system \citep{sudou}, together with a simulation of 
an eccentric evolution of SMBH binaries \citep{iwasawa},
supports this formation scenario through merging of IMBHs. 

\subsection{IMBHs and Gravitational Waves}
In \citet{paperI} (hereafter Paper I), 
we pointed out that gravitational waves from IMBHs can be a trigger to 
prove this process.  If the space-based laser interferometers are in action, then 
their observation ranges ($10^{-4}$--$10$ Hz) are quite reasonable for IMBH mergers.
By accumulating data of merger events, we can specify the IMBH merger scenario
such as they merge hierarchically or monopolistically.  

Later, 
\citet{fregeau2006} discussed the event rates of IMBH--IMBH binary observations
at advanced LIGO 
and VIRGO  
and concluded that we can expect $\sim$ 10 mergers per year. 
This work was followed by
\citet{gair2011}, including the Einstein Telescope project. 
\citet{seoane2010} also discussed the IMBH--IMBH system, including the pre-merger phase.  

\begin{figure}[h]
\plotone{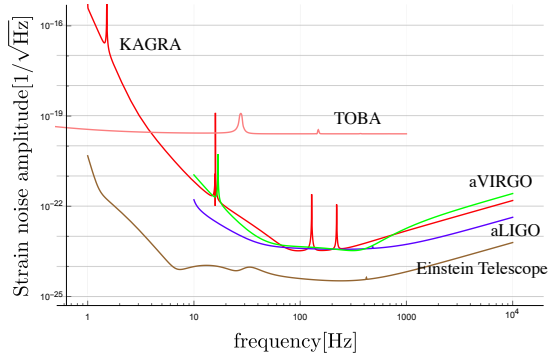}
\caption{Designed strain noise amplitude of the advanced detectors (advanced LIGO, 
advanced VIRGO, and KAGRA) and the planned Einstein Telescope.  We also plotted 
that of a torsion-bar antenna (TOBA). 
\label{fig1-strainnoise}}
\end{figure}

Noting that the second generation of GW interferometers have enough sensitivity 
at 10 Hz and above (see Fig.\ref{fig1-strainnoise}), 
we are able to detect the ring-down gravitational wave of a BH of the mass 
$M < 2\times 10^3 M_\odot $. 

In this article, we therefore discuss how much we can observe BH mergers by finding
their ring-down part using designed ground-based detectors. 
We roughly assume the mass distribution of BHs, $N(M)$, in a galaxy or globular cluster, 
which would be related to the merging history of BHs, and estimate the event rate 
using the designed strain noise of KAGRA, which is at the equivalent level with aLIGO/aVIRGO. 

In addition, recent approaches to gravitational-wave detection using a
torsion-bar antenna (TOBA; \citep{TOBA2010,TOBA2011}) are also quite attractive 
for this purpose 
since it covers the low frequency range (0.1 Hz -- 10 Hz). 
However, the current strain noise amplitude of TOBA is larger compared to those of 
interferometers (see Fig.\ref{fig1-strainnoise}), and we do not discuss the case of TOBA
in this article. 

The organization of the paper is as follows. 
In \S 2, we present the basic equations of gravitational radiation from IMBH binaries.
In \S 3, we estimate the event rate of IMBH mergers under the simplest assumptions on the galaxy distribution and formation process of SMBHs. 
A summary and discussion are presented in \S 4.
Throughout the paper, we use $c$ and $G$ for the
light speed and gravitational constant, respectively.
\section{Black Hole Merger Model}
\subsection{Ring-down Frequency from BHs}\label{sec2}

The gravitational waveform of binary-star mergers which ends up with a single BH, 
has three typical phases: inspiral phase, merging phase, and ring-down phase. 
The waveform in the inspiral phase is called the ``chirp signal" from its feature
of increasing frequency and amplitude.  For the case of GW150914, the frequency was
first caught at 35 Hz, and then it increased to 150 Hz, where the amplitude reached the maximum, which indicates the merger of the binary. The final ``ring-down" signal was 
supposed to be around 300 Hz.

\begin{figure}[h]
\plotone{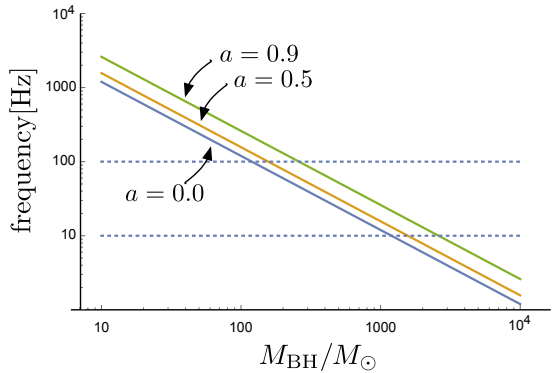}
\caption{Quasi-normal frequency $f_{\rm {qnm}}$ as a function of the mass of BHs $M_T$. If we restrict the observable range is to above 10 Hz for the advanced ground-based interferometers, then the BHs with mass are less than 2000 $M_\odot$ are within the target. 
\label{fig2_qnmKerr}}
\end{figure}

As we mentioned in Paper I, for massive BH binaries with masses greater than 
$10^3 M_\odot$, the inspiral frequencies are less than 1 Hz. 
The wavelength of this frequency range is apparently more than the size of the Earth, 
so that its detection requires interferometers in space.  
On the other hand, the ring-down frequency is simply estimated 
by the quasi-normal frequency of BHs,
$f_R+if_I$, which is determined from the mass and spin of the final BH and is  
estimated to be higher frequency than in its inspiral phase.  
The quasi-normal modes are derived as eigenvalues of the wave equations on the perturbed geometry
(see, e.g. \citet{leaver}). 
For a BH with mass $M_T$ and spin $a$, fitting functions are also known (\citet{echeverria};\citet{BertiCardosoWill2006})
in the form 
\begin{eqnarray}
f_R&=&f_1+f_2(1-a)^{f_3}\\
Q&\equiv&\frac{f_R}{2f_I}=q_1+q_2(1-a)^{q_3}
\end{eqnarray}
where $Q$ is called the quality factor and $f_i, q_i$ are fitting coefficients. 
For the most fundamental mode, which is of the spherical harmonic index $\ell=2$, $m=2$, the
fitting parameters are $f_1=1.5251, f_2=-1.1568, f_3=0.1292, q_1=0.7000, q_2=1.4187,$ and $q_3=-0.4990$
(\citet{BertiCardosoWill2006}).
Recovering the units, we can write the frequency as 
\begin{eqnarray}
f_{\rm {qnm}}&=&\frac{c^3}{2 \pi G M_T} f_R \nonumber \\
&\sim& 3.2\left( \frac{10 M_\odot}{M_T}\right) f_R \mbox{[kHz]}. \label{fqnm}
\end{eqnarray}
We plot $f_{\rm {qnm}}$ in Fig.\ref{fig2_qnmKerr}. 
 
Supposing that advanced GW interferometers can detect $f_{\rm {qnm}}$ above 10 Hz, then 
BHs less than 1200 $M_\odot$ are within the target if BHs are nonrotating ($a=0$), while 
BHs less than $2500 M_\odot$ are in the detectable range for highly rotating cases ($a=0.98$).  

With this simple estimation, we hereafter consider mergers of BHs with total mass less than 2000  $M_\odot$. 

\subsection{Number of Galaxies in the Universe}
In order to model the typical mass of galaxies and its distribution, 
We apply the halo mass function given by \citet{valeostriker}, 
in which they discuss an empirically based, nonparametric model for galaxy luminosities with halo/subhalo masses. 
They apply the Sheth-Tormen mass function \citep{shethtormen} for halo number density, 
\begin{equation}
n_{H}(M)dM=0.322 \left(1+\frac{1}{\nu^{0.6}}\right)\sqrt{\frac{2}{\pi}} \frac{d\nu}{dM}\exp\left(-\frac{\nu^2}{2}\right) dM \label{halomassfunc}
\end{equation}
where $\nu=\sqrt{a} \delta_c (1+z) \sigma(M)$ with $a=0.707$, 
the linear threshold for spherical collapse $\delta_c=1.686$, 
and $\sigma(M)$ is the variance on the mass scale $M$. 
This mass function is roughly $\sim M^{-1.95}$ at low mass. 

\begin{figure}[h]
\plotone{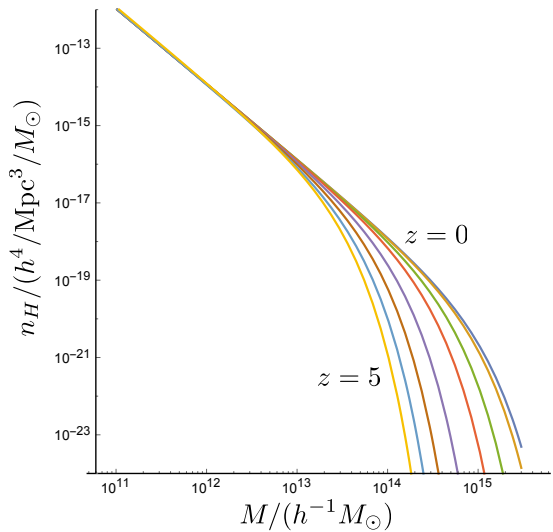}
\caption{Global mass functions for halos (halo and subhalo),  
$n_H(M)$, for $z=0, 0.1, 0.5, 2, 4, 5$ [equation (\ref{halomassfunc})]. 
$n_H(M)$ is in units of $h^4 /{\rm Mpc}^3/M_\odot$.  
$M$ is in units of $h^{-1} M_\odot$, with $h=0.7$. 
\label{fig3_NofM}}
\end{figure}

\citet{valeostriker} also derive an average number of galaxies (subhalos) predicted 
for a parent halo of mass, which is roughly given by $N_{\rm subhalo}\sim M^{0.9}$ (Fig.12 in their paper). 
If we regard this relation as a seed of galaxies, then it 
indicates that a typical galaxy has mass $10^{11}- 10^{12} M_\odot$.

Integrating Equation (\ref{halomassfunc}) by the volume as a function of redshift $z$, 
we can derive the number density of halos (Figure \ref{fig3_NofM}).  
In this process, 
we use the standard cosmology model with current parameters, i.e.   
we use the flat Friedmann model with Hubble constant $H_0$=72 km~s$^{-1}$ Mpc$^{-1}$, matter and dark matter density 
$\Omega_{m0}=0.27$, and 
dark energy (cosmological constant) $\Omega_{d0}=0.73$. 
The luminosity distance $d_L(z)$ is given by 
\begin{eqnarray}
d_L(z)=(1+z) \int_0^z \frac{c \, dz}{H(z)} \label{dL}
\end{eqnarray}
where 
\begin{eqnarray}
H(z)=H_0 \sqrt{(1+z)^3 \Omega_{m0}
+\Omega_{d0}}. 
\end{eqnarray}
The volume of the universe is $V(d)=4 \pi d^3/3$.

\begin{figure}[h]
\plotone{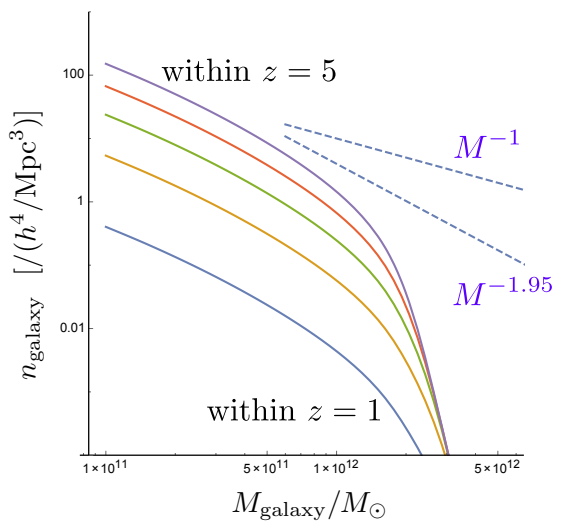}
\caption{Number density of galaxies, $n_{\rm galaxy}(M)$. 
\label{fig4_Ngalaxy}}
\end{figure}

Combining these two functions (average number of galaxies and the number density of halos), we get the number density of galaxies $n_{\rm galaxy}(M,z)$, which we show in 
Fig.\ref{fig4_Ngalaxy}.  If we integrate it by $M$ and $z$ as  
\begin{equation}
N_{\rm galaxy} (z)=\int^z_0 dz \int_{M_1}^{M_2} n_{\rm galaxy}(M, z) dM, 
\end{equation} 
then we get the number of galaxies.  We set $M_1=10^9M_\odot$ and $M_2=10^{13}M_\odot$.

From the recent ultraviolet luminosity density of star forming galaxies, 
star formation rate density $\rho_{\rm SFR}(z)$ is fit as 
\begin{eqnarray}
\rho_{\rm SFRp}(z) &=&
\frac{0.009+0.27 (z/3.7)^{2.5}}{1+(z/3.7)^{7.4}}+10^{-3}\\
\rho_{\rm SFRr}(z) &=&
\frac{0.009+0.27 (z/3.4)^{2.5}}{1+(z/3.4)^{8.3}}+10^{-4}
\end{eqnarray}
for metal poor stars and metal rich stars, respectively \citep{RobertsonEllis2010, RobertsonEllis2012}. 
If we sum these two (normalized $\rho_{\rm SFRp}$ and normalized $\rho_{\rm SFRr}$) evenly, the peak location 
is at $z=3.26$.  We then obtain 
\begin{equation}
N_{\rm galaxy} (z)=\int^z_0 \rho_{\rm SFR}(z) dz \int_{M_1}^{M_2} n_{\rm galaxy}(M, z) dM. 
\label{eq.Ngalaxy}
\end{equation}
The typical numbers of our model are shown in Table.\ref{table.galaxymodel}.
These numbers are slightly larger than the latest observation by \cite{Conselice2016}, but 
our model produces the same order and its evolution history for $N_{\rm galaxy}$ as theirs. 

\input{table1}
\subsection{Number of BHs in a Galaxy}
We next estimate the number of BH candidates in a galaxy. 
Recently, \citet{inutsuka} developed a scenario of galactic-scale star formation
from a giant molecular cloud.  Their model includes both the growth of molecular clouds
and the destruction of magnetized molecular clouds by radiation.  Simulations and steady-state analysis 
show that the mass density function of molecular clouds, $n_{\rm cl}(M_g)$, converges at the 
Schechter-like function, 
\begin{equation}
n_{\rm cl}(M_{\rm cl})\sim M_{\rm cl}^{-1.7} \exp \left( - \frac{M_{\rm cl}}{M_{\rm cut}} \right) 
\label{GMCdensity}
\end{equation}
where the cutoff mass $M_{\rm cut}=10^6 M_\odot$. 

On the other hand, many $N$-body simulations report that there is a simple relation between 
the mass of the most massive cluster $m_{\rm max}$ and the total mass of the 
molecular cloud $M_{\rm cl}$, 
\begin{equation}
m_{\rm max}=0.20 M_{\rm cl}^{0.76}.  \label{eq.McMcmax}
\end{equation}
The single-line fit can be seen for the wide range $M_{\rm cl}/M_\odot = 10^0 - 10^7$ 
(see Fig.6 in \cite{Fujii2015}). 

\begin{figure}[h]
\plotone{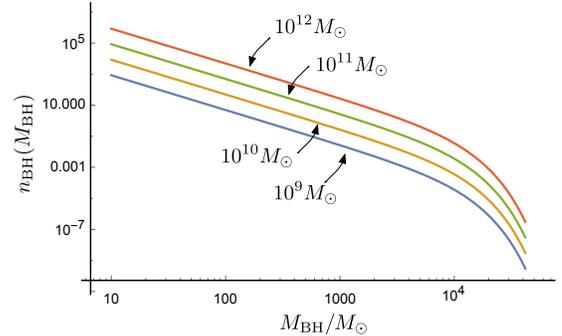}
\caption{Number density of BHs { per galaxy} as a function of BH mass for different total mass of galaxies 
$M_{\rm galaxy}=10^9 M_\odot, \cdots, 10^{12} M_\odot$. 
\label{fig5_buildingblockBH}}
\end{figure}

We therefore combine these results, and we suppose that each molecular cloud forms a single BH 
in its core if it is more than $10 M_\odot$, and we suppose that these BHs become ``building blocks" for 
forming stellar-sized and intermediate-mass BHs. 
{ Many $N$-body simulations suggest that massive objects will accumulate in the center of a galaxy owing to  dynamical friction, so that we modeled that these seed BHs accumulate and merge repeatedly (as we model below), resulting in IMBHs and SMBHs.  We do not specify where these mergers occur, but we count our BH mergers after we set up the initial seeds. }
We show the number density of BHs { in a galaxy}, $n_{\rm BH}(M_{\rm BH})$ in Fig. \ref{fig5_buildingblockBH}.

\subsection{Number of BH Mergers in a Galaxy}
In Paper I, we considered two toy models for 
formation of SMBHs: hierarchical growth and runaway growth.
The hierarchical growth model is the case in which 
two nearby equal-mass BHs merge simultaneously and continue their mergers. 
The runaway growth model is, conversely, where only one BH grows 
itself by continual mergers with surrounding BH companions. 

The recent $N$-body simulations report that the hierarchical merger process is plausible both 
for the massive clusters ($10^4$-$10^6 M_\odot$; 
see e.g. \citet{Fujii2015}) and for stellar-mass BHs (see e.g. \cite{FujiiTanikawaMakino2016}). 

We therefore simply assume that BHs formed at cores of clouds will accumulate each other hierarchically, 
i.e. the mass and the number of BHs at steps from $k$ to $k+1$ can be expressed simply by  
\begin{eqnarray}
M_{k+1}&=&2 M_{k}, \\
N_{k+1}&=& N_{k}/2. 
\end{eqnarray}
The mass of a BH merger, then, obeys the distribution $M^{-1}$
(see footnote \footnote{
Suppose we have a cluster of the total mass $M_c$ that consists
of $N_0$ equal-mass BHs. 
This means that each BH mass is initially  
${M_c/N_0}$. They continue to form binaries and merge together, 
which indicates that there are $N_0/2^{i-1}$ binaries 
for the $i$th generation that forms BHs with the masses $M=2^{i-1} M_c/N_0$.
The model shows only the discrete distribution of the BH mass, but the number of binaries 
$N(M)$ 
can be approximated with the number of initial fractions in a cluster, 
$N(M)=M_c/M$. 
}). 

On the other hand, we know empirically that the mass of the central BH of the galaxy, 
$M_{\rm SMBH}$, and the total mass of the galaxy, $M_{\rm galaxy}$, 
has a relation
\begin{equation}
M_{\rm SMBH}=2\times 10^{-4} M_{\rm galaxy} 
\end{equation}
(or equal to $10^{-3}$ of the bulge mass; see, e.g. \cite{King2003},\cite{McConnellMa}). 

\begin{figure}[h]
\plotone{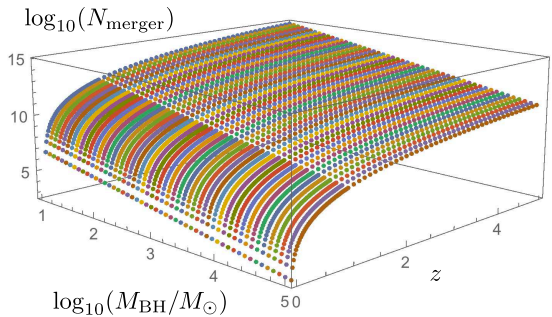}
\caption{Cumulative distribution function of { the number of 
BH mergers $N_{\rm merger} (M_{\rm BH})$ as a function of the redshift $z$. 
We express number with binned one, of which we binned 20 for one order in $M_{\rm BH}$. }
\label{fig6_Nmerger}}
\end{figure}

Combining these facts, for a certain galaxy with $M_{\rm galaxy}$, we pick up BHs with total mass 
$M_{\rm SMBH}$ (equation above), obeying the mass distribution of Fig.\ref{fig5_buildingblockBH}.
We suppose that picked-up BHs will form an SMBH in its series of mergers in the hierarchical model. 
Together with galaxy distribution function $n_{\rm galaxy}(M,z)$, we are able to count the possible 
events of BH mergers, $N_{\rm merger}(M_{\rm BH}, z)$, in the universe, which we show in Fig.\ref{fig6_Nmerger}. 

In the next section, we further take into account the detectors' detectable distance $D(M, a, \rho)$ (with BH spin parameter $a$, energy emission rate of merger, S/N $\rho$). 
In \S 4, we estimate the observable event rate, 
\begin{equation}
{\rm Event~Rate}~R {\rm [/yr]}=\frac{N_{\rm merger}(z)}{V(D/2.26)},  
\label{eq.eventratedef}
\end{equation}
where the factor 2.26 is for averaging the distance for all directions (\cite{FinnChernoff}). 

\section{Signal-to-noise Ratio and Detectable Distance}
\subsection{S/N}
Let the true signal $h(t)$, the function of time, be detected as
a signal, $s(t)$, which also includes the unknown noise, $n(t)$:
\begin{equation}
s(t)=h(t)+n(t). 
\end{equation} 
The standard procedure for the detection is judged by the optimal S/N 
ratio (SNR), $\rho$, which is given by 
\begin{equation}
\rho=2 \left[ \int_0^\infty \frac{\tilde{h}(f)\, \tilde{h}^*(f)}{S_n(f)} df \right]^{1/2}, 
\end{equation}
where $\tilde{h}(f)$ is the Fourier-transformed quantity of the wave, 
\begin{equation}
\label{fourier}
{\tilde h}(f) = \int_{-\infty}^{\infty} e^{2 \pi i f t}\, h(t) \, dt,
\end{equation}
and $S_n(f)$ is the (one-sided) power spectral density of 
strain noise of the detector, as we showed in Fig. \ref{fig1-strainnoise}.
In this paper, for KAGRA (bKAGRA), we use a fitted function 
\begin{equation}
\sqrt{S_n(f)}=10^{-26} \left( \frac{6.5\times 10^{10}}{f^8} + \frac{6\times 10^6}{f^{2.3}}+1.5f \right), 
\end{equation}
where $f$ is measured in Hz, 
as was used in \cite{NakanoTanakaNakamura2015}. 

\subsection{S/R of Ring-down Wave}
For the ring-down gravitational wave in the presence of a BH, 
the waveform is modeled as 
\begin{equation}
h(t) = A \cos(2 \pi f_R (t-t_0) + \psi_0) e^{-(t-t_0)/\tau}
\label{ringdownbasic}
\end{equation}
where $f_R$ is the oscillation frequency, and $\tau$ is the decaying time constant, and 
$t_0$ and $\psi_0$ are the initial time and its phase, respectively (we simply set $t_0=\psi_0=0$). 
The parameter $\tau$ is normally expressed using a 
quality factor, $Q\equiv  \pi {f_R} \tau$, or $f_I = 1/(2\pi \tau)$. 
The waveform, Equation (\ref{ringdownbasic}), is then written as 
\begin{equation}
h(t)\sim A e^{i 2\pi (f_R+if_I) t}
\end{equation}
where we call $f_R+if_I$ the quasi-normal frequency, which is obtained 
from the perturbation analysis of BHs, and its fitting equations are shown in Equation (\ref{fqnm}).

Following \citet{flanagan1998}, we use the energy spectrum formula for 
the ring-down wave  
\begin{eqnarray}
\frac{dE}{df} &=& \frac{{A}^2M^2 f^2}{32\pi^3 \, \tau^2} \nonumber \\&&
\times \bigg\{
\frac{1}{\left[(f-{f_R})^2 +f_I^2\right]^2} + 
\frac{1}{\left[(f+{f_R})^2 +f_I^2\right]^2} \bigg\}
\label{dEdfqnr} \nonumber \\
&\approx& \frac{1}{8} {A}^2  M^2 f_R Q \, \, \delta(f -
f_R) \left[1 + O(1/Q)\right]. \nonumber \\&&
\label{dEdfqnrapprox}
\end{eqnarray}
where $M$ is the total mass of the binary, $M=m_1+m_2$.
We then obtain 
\begin{eqnarray}
E_{\rm ringdown} &\approx& \frac{1}{8} {A}^2 M^2 f_R Q. 
\label{qnrenergytot}
\end{eqnarray}

\begin{figure}[h]
\plotone{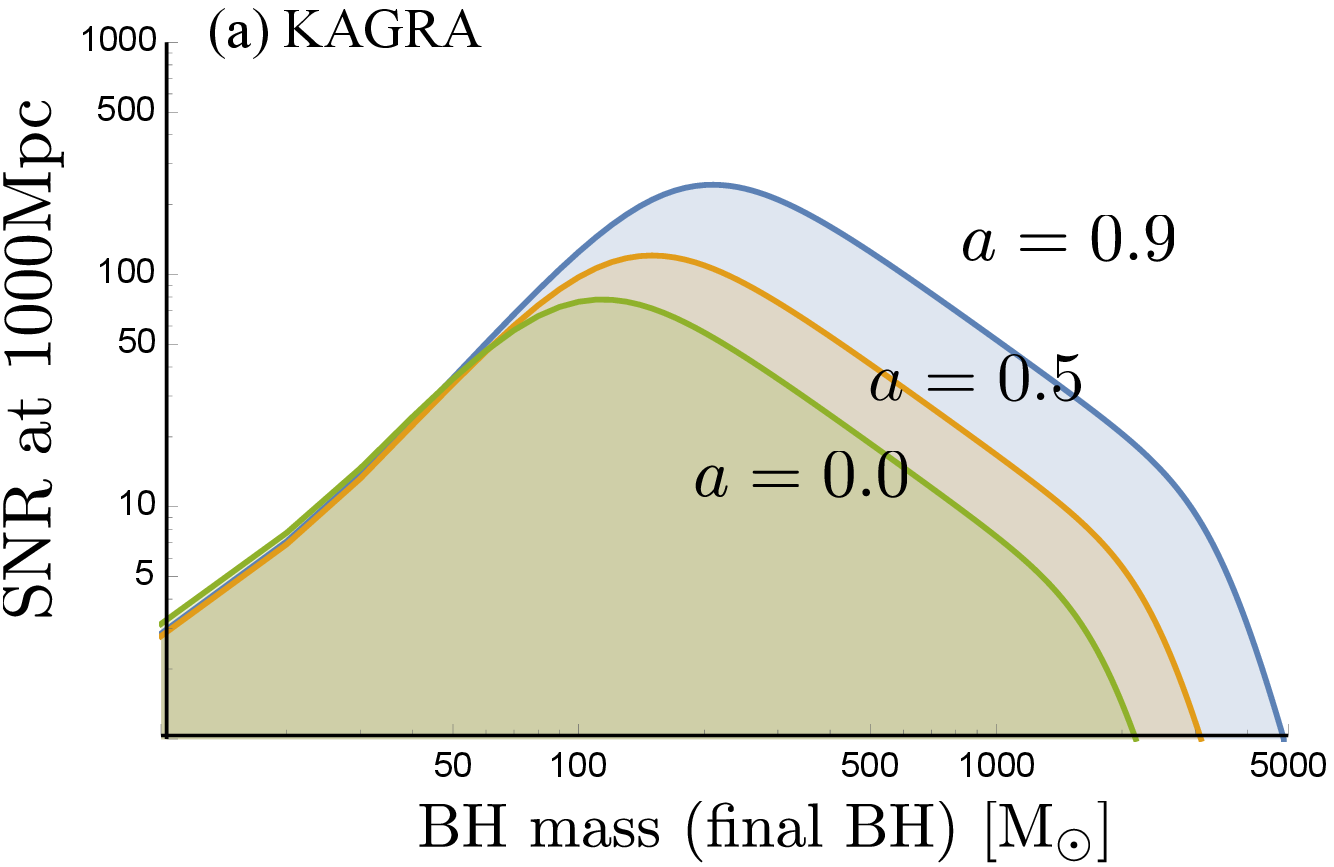}
~\\
\plotone{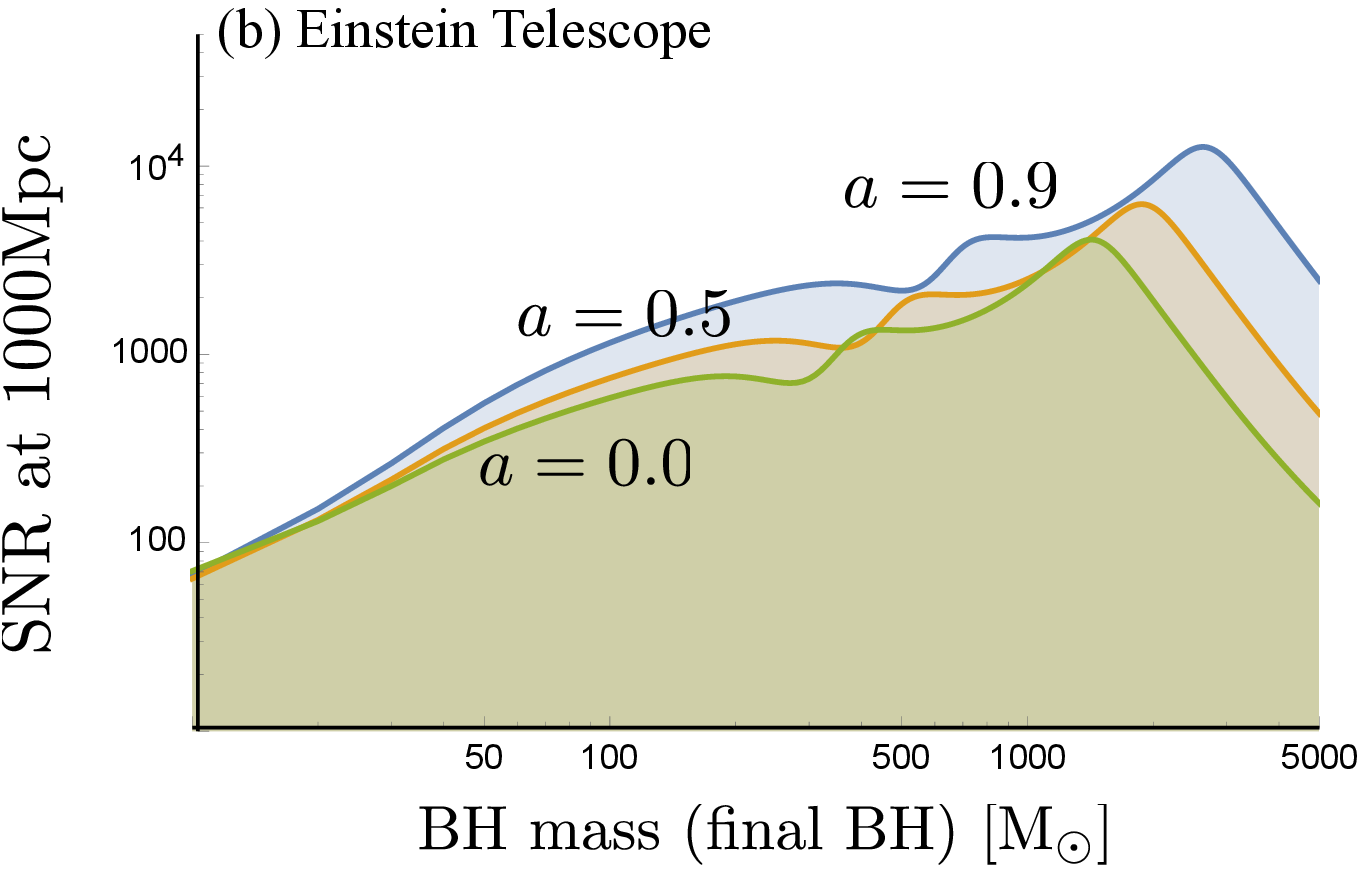}
~\\
\plotone{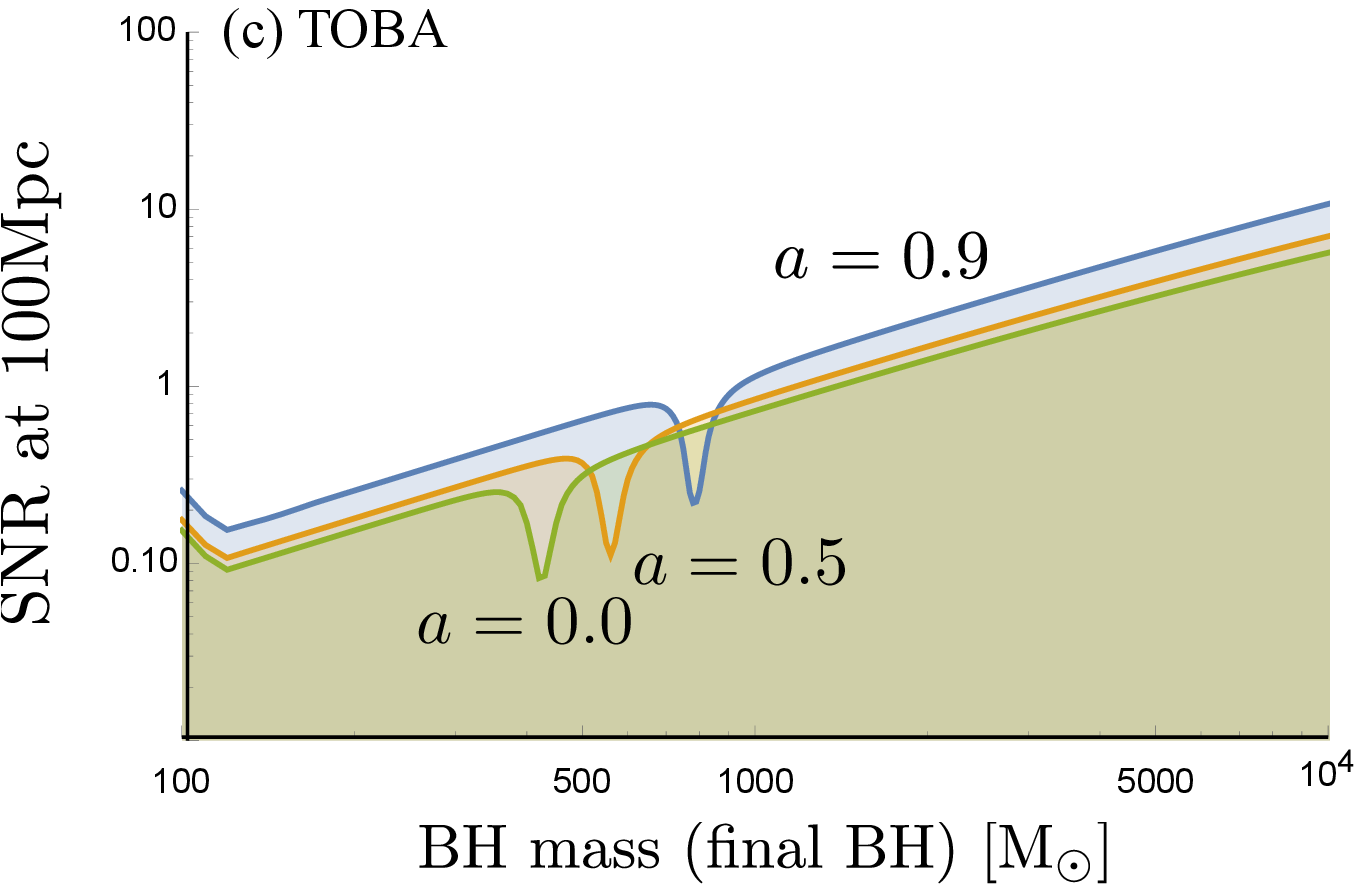}
\caption{S/R for ring-down waves from a BH with spin parameter $a$, which appears 
at a distance of 1 Gpc.  Panels (a) and (b) are for KAGRA and the Einstein Telescope, respectively. 
We see that ring-down frequencies of IMBHs (especially for 100--400 $M_\odot$) are the best target for both KAGRA and the Einstein Telescope.  Panel (c) is for TOBA, and the distance is
estimated at 100 Mpc. 
\label{fig7_SNR}}
\end{figure}

Let 
$\epsilon_r(a)\equiv E_{\rm ringdown}/M$, which expresses 
the energy fraction of the emitted gravitational wave to the total mass. 
As we cited in the introduction, GW150914 and GW151226 show us $a=0.67$ and  $0.74$ and energy 
emission rate $4.6\%$ and $ 4.1\%$ of the total mass, respectively. 
The associated numerical simulation of GW150914 (SXS:BBH:0305) \footnote{SXS Gravitational Waveform Database  
(https://www.black-holes.org/waveforms/)} 
shows that the $4.0\%$ of the total mass is emitted before the merger
\footnote{We thank Hiroyuki Nakano for pointing out this ratio.}. 
That is, the ring-down part emits the energy around 0.6 \% of the total mass.
If we use ${A}\sim 0.4$, then we recover the ratio $\epsilon_r(0.67)=0.58\%$
(it also produces, e.g. $\epsilon_r(0.0)=0.236\%$, $\epsilon_r(0.5)=0.425\%$, $\epsilon_r(0.9)=1.23\%$, 
$\epsilon_r(0.98)=2.98\%$). 
The magnitude of this $A$ is also consistent with the quadrupole formula. 

The S/N is, then, expressed using the inertial mass $\mu=m_1m_2/M$ and 
the redshift of the source $z$, 
\begin{eqnarray}
&&\rho^2 =
\frac{8}{5} \, \frac{\epsilon_r(a) }{f_R^2} \,
\frac{ (1+z) M }{S_h(f_R/ (1+z))} \,
\nonumber \\
&& ~~~~~~ \times
\left( \frac{(1+z) M }{d_L(z) }\right)^2 \,
\left( \frac{4 \mu }{ M }\right)^2. 
\label{qnrsnrII}
\end{eqnarray}
Up to here, we see that the S/N  is larger when the BH spin $a$ is large, and it 
reaches a maximum when $m_1=m_2$.

In Fig \ref{fig7_SNR}, we plot the S/N of ring-down waves from a BH  
at a distance of 1 Gpc at KAGRA for $A=0.4$. 
The results depend on the BH spin parameter $a$, but 
we see that ring-down frequencies of IMBHs (especially for 100--400 $M_\odot$) 
are the best target for both KAGRA and the Einstein Telescope.  

\subsection{Detectable Distance}
By specifying the BH mass and spin, together with $\rho$, 
we can then find the distance $d_L$ that satisfies eq. (\ref{qnrsnrII}). 
We call this distance the detectable distance, $D(M,a,\rho)$.

\begin{figure}[h]
\plotone{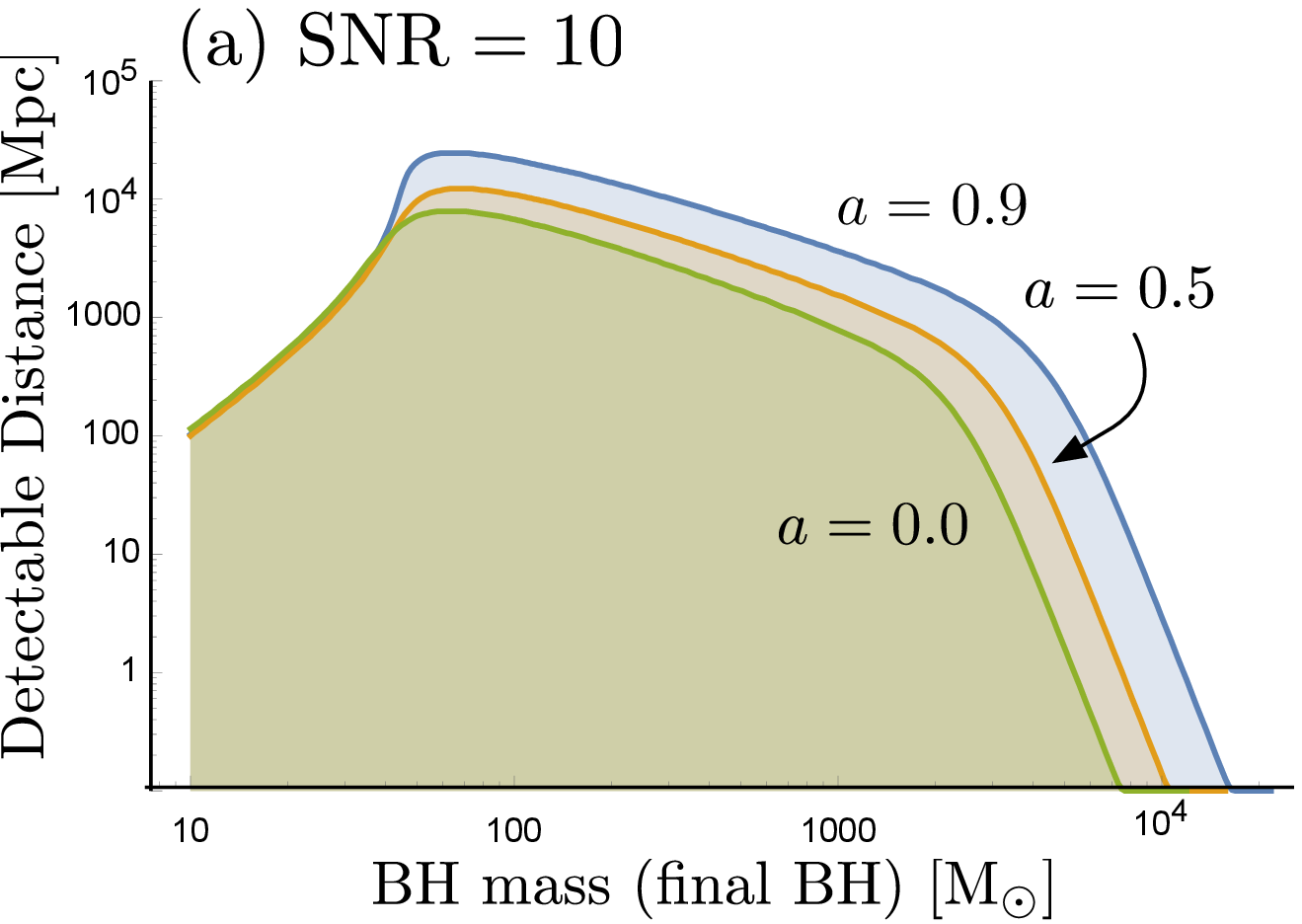}\\
~\\
\plotone{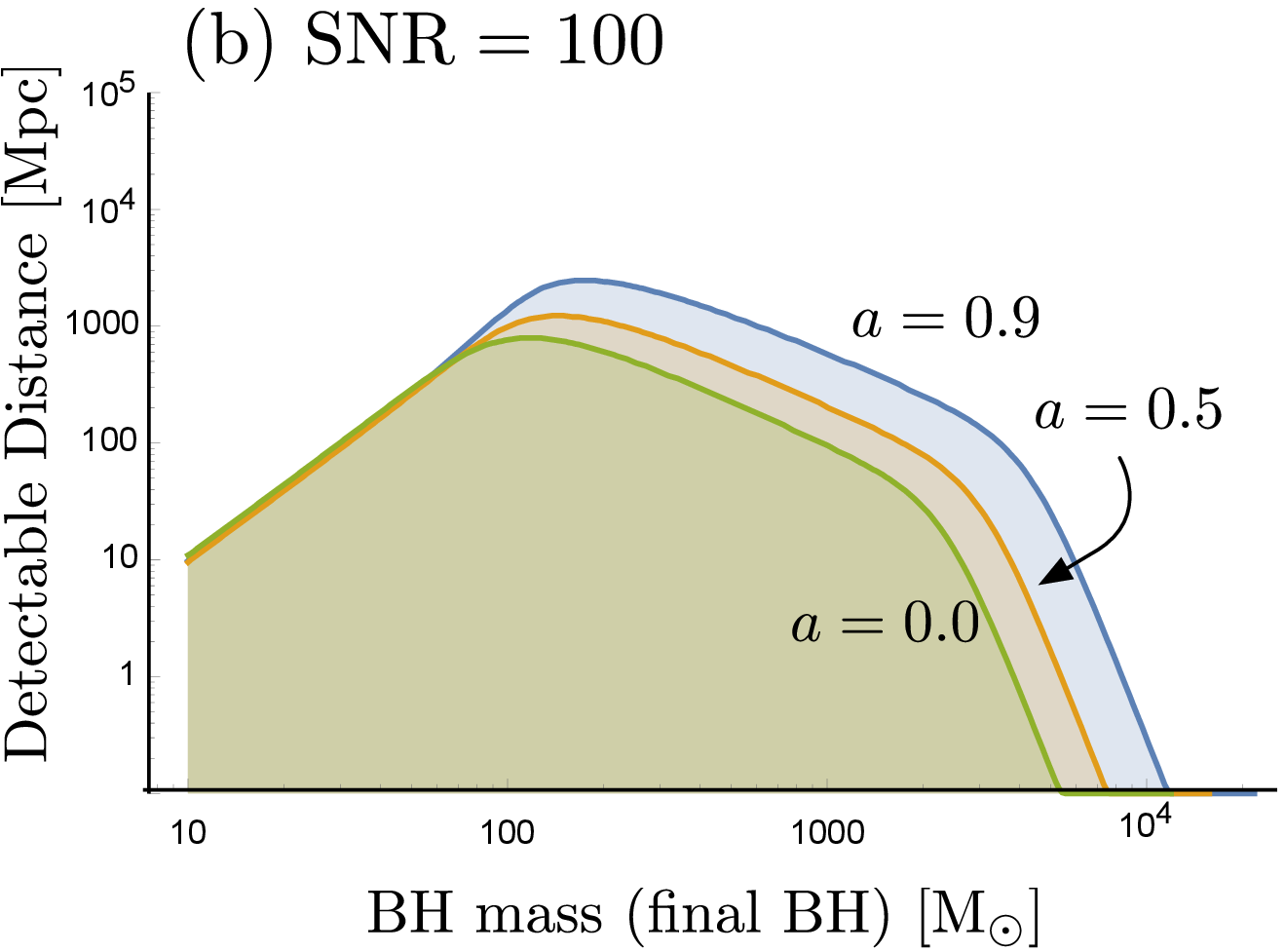}
\caption{Detectable distance $D$ of the ring-down signal at KAGRA. 
S/R is set to (a) 10 and (b) 100.  
\label{fig8_distance}}
\end{figure}

We converted Fig.\ref{fig7_SNR} into the plots of detectable distance $D$ 
with a function of BH mass $M$ for S/R=10 and 100.  
We show them in Fig.\ref{fig8_distance} for KAGRA.
We see that the designed KAGRA covers at least 100 (10) Mpc at S/R=10 (100) for $10M_\odot <M $, 
and KAGRA covers 1 Gpc at S/R=10 for $ 40M_\odot <M < 1000 M_\odot$. 

\section{Event Rate}
\begin{figure*}[h]
\plottwo{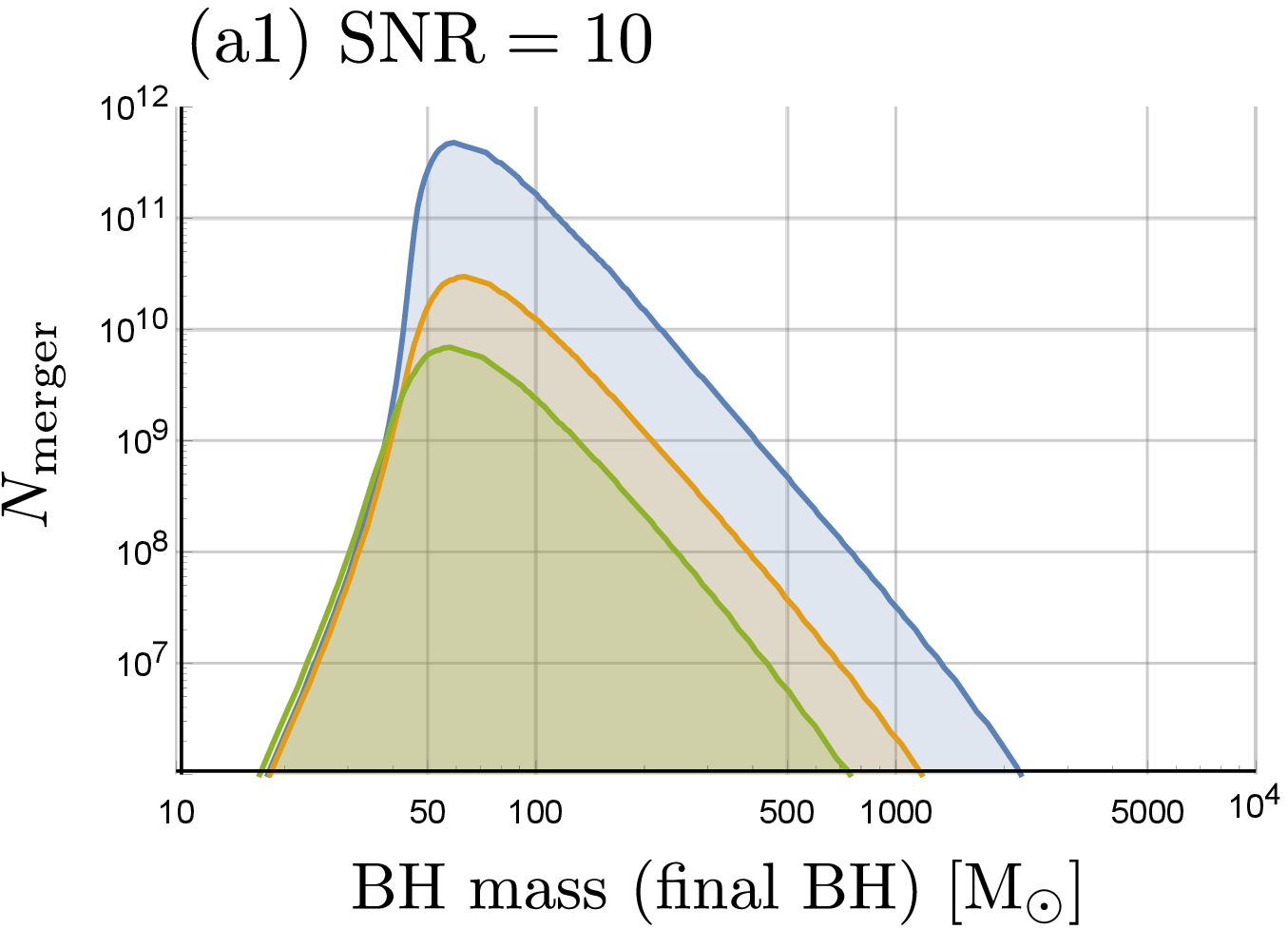}{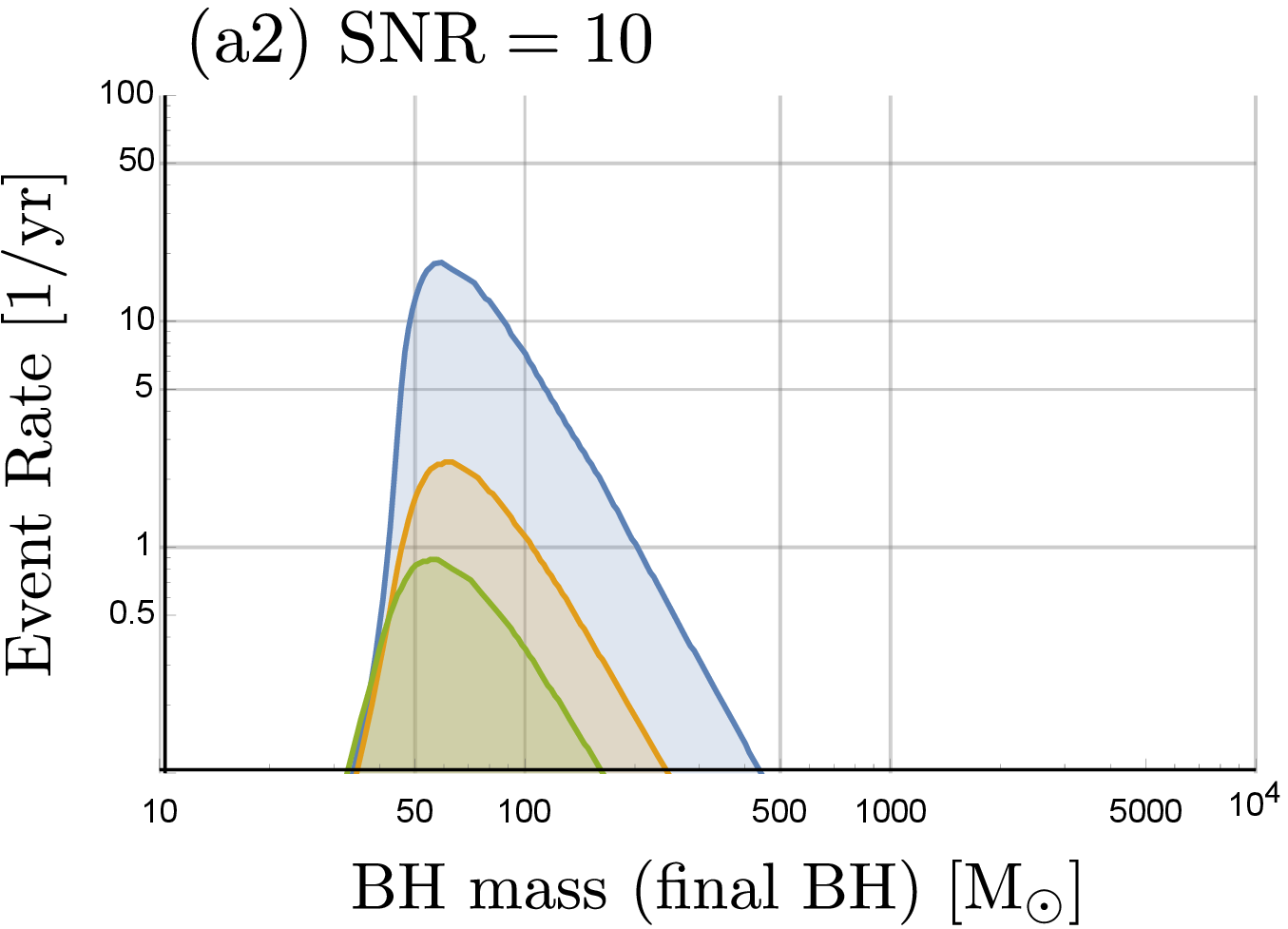}\\~\\
\plottwo{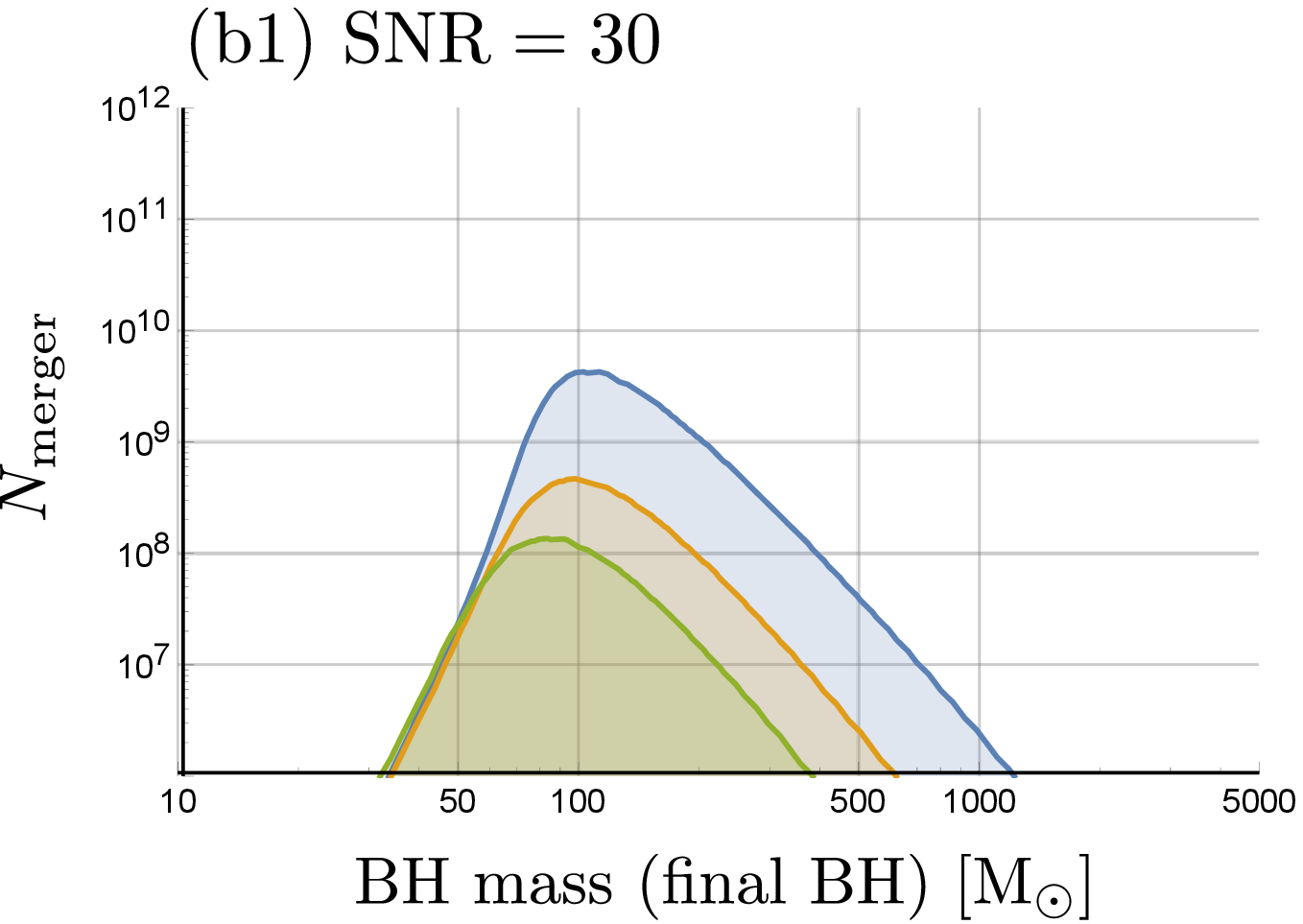}{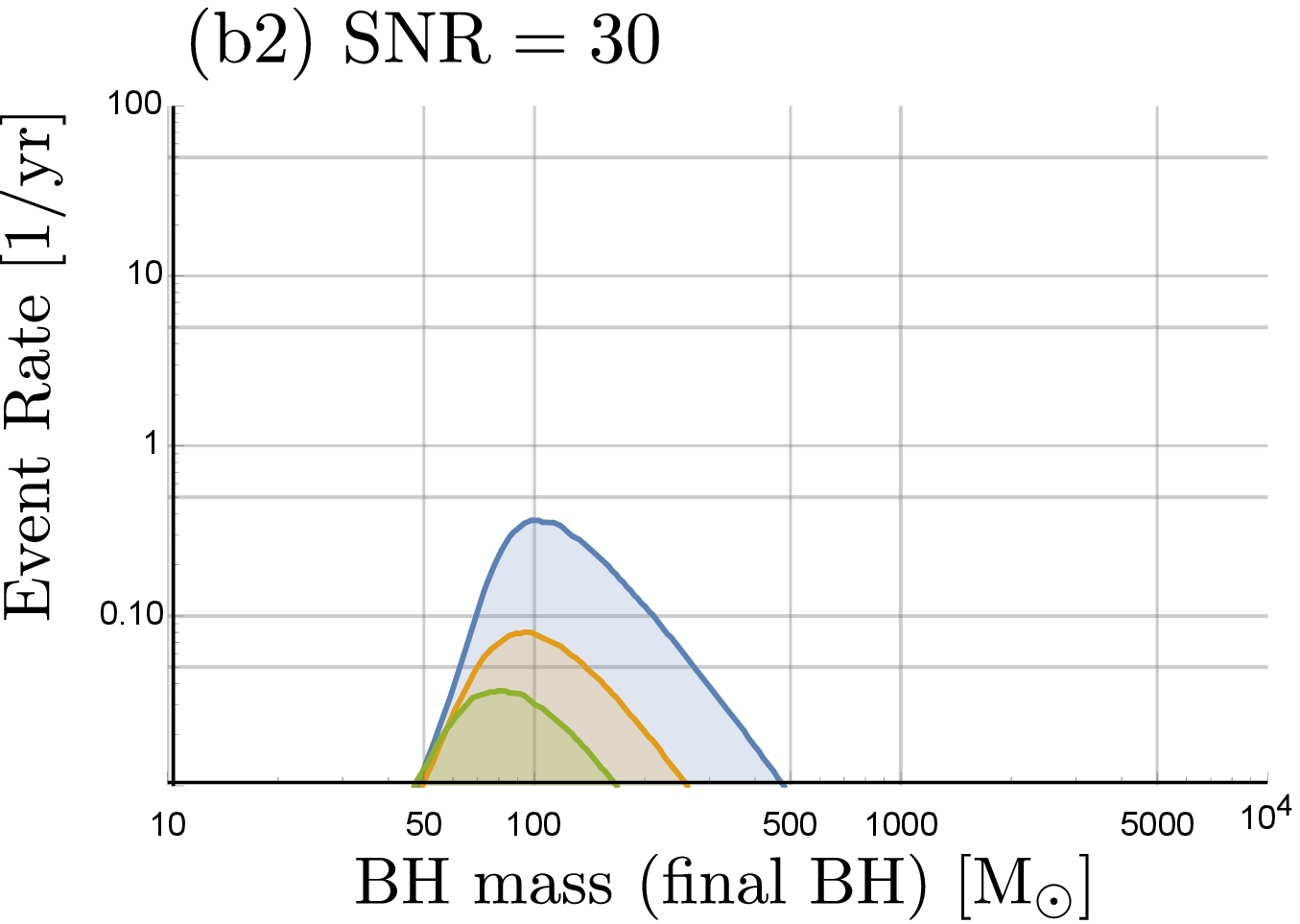}
\caption{{ Number of BH mergers within the detectable distances (a1, b1) and event rate $R$ (a2, b2) } as a function of BH mass $M$ with S/N ratio $\rho=10$ and $30$ for KAGRA. 
Three distributions for each figure are of $a=0.9, 0.5, $ and $0.0$ (from largest to lowest), respectively. 
\label{fig9_eventrateKAGRA}}
\end{figure*}

Using the detectable distance $D(M,a,\rho)$ obtained in the previous section, 
we set the upper limit of $z$ for integrating eq. (\ref{eq.Ngalaxy}) to obtain the number of galaxies, $N_{\rm galaxy}$, 
and then obtain the number of BH mergers, $N_{\rm merger}$, according to the procedure shown in \S 2. 
We show $N_{\rm merger}$ in Figures \ref{fig9_eventrateKAGRA} (a1) and (b1) for S/R=10 and 30, respectively. 

The event rate $R$, then, is estimated by eq. (\ref{eq.eventratedef}). 
We show them in Figures \ref{fig9_eventrateKAGRA} (a2) and (b2). 
Previous works (e.g., \cite{miller2002,will2004}) assume that the number of events to the merger sources is roughly $\sim 10^{-10}$, 
which can be seen in our Figures \ref{fig9_eventrateKAGRA} (a1) and (a2) for higher-spinning BH cases. 

\begin{figure}[h]
\plotone{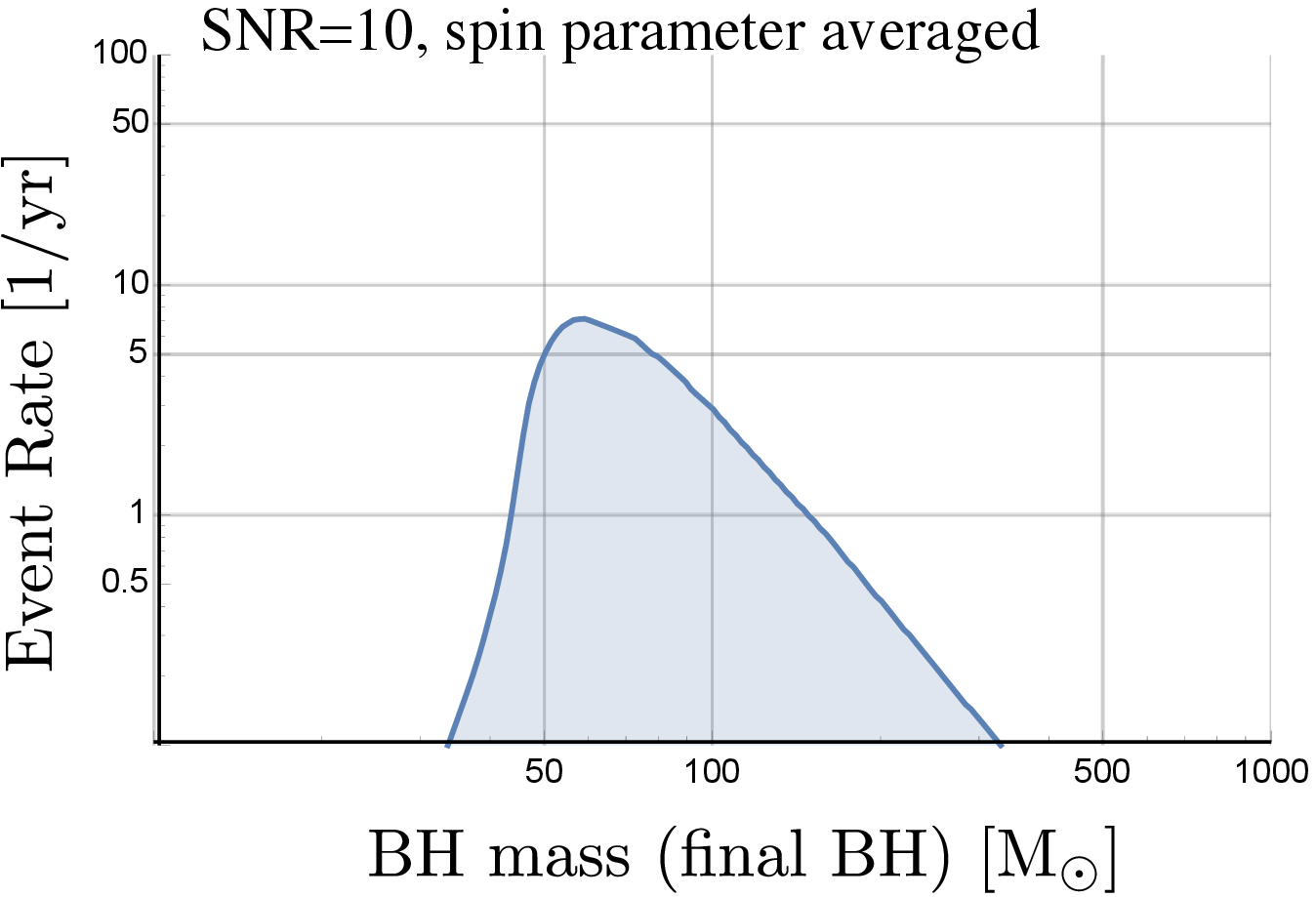}
\caption{Event rate $R$ as a function of BH mass $M$ with S/N ratio $\rho=10$ for KAGRA. 
Spin parameter dependences are averaged. 
\label{fig10_eventrateKAGRAave}}
\end{figure}

Fig. \ref{fig9_eventrateKAGRA} is for specifying the BH spin parameter $a$, but if we assume that $a$ is homogeneously distributed, then 
the averaged $R$ is estimated as in Figure \ref{fig10_eventrateKAGRAave}. 

The event rate versus mass distribution of Fig. \ref{fig10_eventrateKAGRAave} has its peak $R\sim 7.13$[/yr] 
at $M\sim 59.1 M_\odot$ ($200$--$375$  [Hz]  for $a=0$--$0.9$).  
{ It is interesting to find out that this peak mass matches with the final BH mass of GW150914. }
The mergers of the range above $R>1$ [/yr] have mass $40M_\odot < M < 150 M_\odot$. 
The total number of events above $R>1$ [/yr]  is $\sim 211$. 

{ 
Our event rate sounds similar to that of other groups. For example,  
the LIGO-Virgo group updated their estimated event rates after the detection of GW150914 
as 2--600 Gpc$^3$ yr$^{-1}$ assuming BH mass distribution models
as flat or power law ($\sim M^{-2.35}$; \cite{LIGOeventrate}). 
\citet{Kinugawa2014} estimate as 70--140 yr$^{-1}$ from their Population III model.
\citet{Inoue2016} estimate as $<60$ yr$^{-1}$ from their BH merger model inferred 
from the luminosity function of ultraluminous X-ray sources.
However, our model predicts BH mergers with $M>100M_\odot$, which will be a key to test our model in the future. 
}
\section{Summary}
Based on a bottom-up formation model of an SMBH via IMBHs, we estimate the expected observational profile of gravitational waves at ground-based detectors.
{ We simply modeled that cores of molecular clouds become BHs if they are more than 10 $M_\odot$, which 
become building blocks for forming larger BHs. We also modeled that BH mergers are accumulations of equal-mass ones and suppose that these occur hierarchically.  We did not include gas accretion after a BH is formed. }

At the designed KAGRA (or equivalent advanced LIGO/VIRGO), with the most standard criterion of the S/N  $\rho=10$, we find that the mass distribution of BH mergers has its peak at $M\sim 60M_\odot$, and we can detect also BHs in the range $40M_\odot < M < 150 M_\odot$ in a certain event rates. 

{ 
Detailed numbers depend, of course, depend on model settings and model parameters. 
We assume that all the galaxies in the universe evolve in the single scenario, which will overestimate the event rate if some SMBHs are formed from the direct collapse of gas clouds.  We also ignore galaxy mergers, which are another route of forming SMBHs.  These issues will lower the merger event rates, so that 
our event rates can be understood at the maximum number.  However, the profiles of event rates in terms of BH mass (Fig.\ref{fig10_eventrateKAGRAave}) will remain the same; therefore, our model's feature, the existence of the gravitational-wave events with BHs larger than 100 $M_\odot$, 
will be tested by accumulating actual events. }

We conclude that the statistics of the signals will give us both a galaxy distribution and a formation model of SMBHs, as well as in the future cosmological models/gravitational theories.

\section*{Acknowledgments}
{ We thank the anonymous referee for constructive suggestions. }
This work was supported in part by the Grant-in-Aid for
Scientific Research Fund of the JSPS (C) No. 25400277 (H.S.), and also by 
MEXT Grant-in-Aid for Scientific Research on Innovative Areas ``New Developments 
in Astrophysics Through Multi-Messenger Observations of Gravitational Wave Sources" (No. 24103005)
(N.K.).

\end{document}

%% file: table1.tex
\begin{table}[h]
\caption{\label{table.galaxymodel} Typical numbers of our galaxy model:
the number of Galaxies $N_{\rm galaxy} (z)$, eq. (\ref{eq.Ngalaxy}), and 
the number density of Galaxies $n_{\rm galaxy}$. }
\begin{tabular}{c|cll}
\hline
$z$&$N_{\rm galaxy} (z)$&\multicolumn{2}{c}{$n_{\rm galaxy}$} \\
\hline
1&$1.18\times 10^{9}$& $1.0 \times 10^{-3}$/Mpc$^3$ &for $z<1$\\
2&$9.45\times 10^{10}$& $6.5 \times 10^{-3}$/Mpc$^3$&for $1<z<2$\\
3&$5.23\times 10^{12}$& $2.4 \times 10^{-2}$/Mpc$^3$&for $2<z<3$\\
\hline
\end{tabular}
\end{table}